\documentclass[a4paper,11pt]{article}
\pdfoutput=1

\usepackage{jheppub} 

\usepackage{subfig}

\newlength{\matrixheight}

\title{\boldmath Optimisation of the vertex detector and measurement of Higgs decays to second-generation quarks at the CEPC}

\author[a,b]{Jialin Li,}
\author[c]{Hao Liang,}
\author[d,h]{Kaili Zhang,}
\author[f]{Yifan Zhu,}
\author[a,b,*]{Jun Guo,}
\author[b,g]{Haijun Yang,}
\author[d,e,*]{Manqi Ruan}

\affiliation[a]{National Key Laboratory of Dark Matter Physics, School of Physics and Astronomy, Shanghai Jiao Tong University, Shanghai 200240, China}

\affiliation[b]{Key Laboratory for Particle Astrophysics and Cosmology (KLPPAC-MoE), Shanghai Key Laboratory for Particle Physics and Cosmology (SKLPPC), Shanghai 200240, China}

\affiliation[c]{Laboratoire Leprince-Ringuet, CNRS, Ecole polytechnique, Institut Polytechnique de Paris, \\ 91120 Palaiseau, France.}

\affiliation[d]{Institute of High Energy Physics, Chinese Academy of Sciences, Beijing 100049, China}

\affiliation[e]{School of Physical Sciences, University of Chinese Academy of Sciences (UCAS), Beijing 100049, China}

\affiliation[f]{National Key Laboratory of Dark Matter Physics, Tsung-Dao Lee Institute and School of Physics and Astronomy, Shanghai Jiao Tong University, Shanghai 201210, China}

\affiliation[g]{National Key Laboratory of Dark Matter Physics, School of Physics and Astronomy and Tsung-Dao Lee Institute, Shanghai Jiao Tong University, Shanghai 200240, China}

\affiliation[h]{China Spallation Neutron Source Science Center, Dongguan 523803, China}

\emailAdd{jun.guo@sjtu.edu.cn}  
\emailAdd{ruanmq@ihep.ac.cn}  
\note[*]{Corresponding authors.}

\abstract{
The vertex detector is crucial for precision measurements of the Higgs boson at the electron-positron Higgs factory. Benchmarked with $H \to c\bar{c}$ and $H \to s\bar{s}$ measurements in the $\nu\bar{\nu}H$ channel, we perform an optimisation study on the inner radius and spatial resolution of the vertex detector using the Jet Origin Identification (JOI) framework, which determines the parton flavour of jets using advanced Artificial Intelligence (AI) algorithm. 
We observe that, compared to the reference detector configuration, halving the inner radius and spatial resolution improves the transverse and longitudinal impact parameter resolution approximately by a factor of two, while increasing the accuracy and significance of the $H \to c\bar{c}/s\bar{s}$ measurement by 4\% and 8\%, respectively. Conversely, doubling these parameters results in comparable degradation, with variations in the inner radius being the dominant factor. Our results provide guidance for detector design and highlight promising prospects for identifying the $H \to s\bar{s}$ decay mode at future Higgs factories.
}

\keywords{$e^{+}e^{-}$ Experiments, Higgs decay, Jets identification}

\begin{document}
\maketitle
\flushbottom

\section{Introduction}


Electron–positron colliders provide uniquely clean environments that facilitate precision studies of the Higgs boson properties, crucial for probing physics beyond the Standard Model (SM)~\cite{ES}. 
Among proposed future facilities, the Circular Electron–Positron Collider (CEPC) is designed with a large circumference of 100~km and two interaction points~\cite{CEPC_CDR_Acc}, and will operate at various centre-of-mass energies. According to its Technical Design Report (TDR)~\cite{CEPC_TDR_Acc,snowmass}, the CEPC is planned to accumulate an integrated luminosity of 20~ab$^{-1}$ at 240~GeV as a Higgs factory~\cite{an2019precision}, yielding $4\times10^6$ Higgs bosons and enabling unparalleled precision in Higgs coupling measurements. In addition, it is capable of running at the Z pole (91.2~GeV), the $W^+W^-$ threshold (160~GeV), and could potentially be upgraded to reach the $t\bar{t}$ threshold at 360~GeV.
 
Measurements of the $H\to c\bar{c}$ and $H\to s\bar{s}$ directly test the Yukawa couplings between the Higgs and the second-generation quarks. The SM predicts the \(H\to c\bar{c}/s\bar{s}\) has a branching ratio of 0.029/$2.3\times10^{-4}$~\cite{SM_BR,Hss_BR,PDG}. 
Notably, these correspond to about 120k $H\to c\bar{c}$ events and about 1k $H\to s\bar{s}$ events at the CEPC when considering all Higgs production modes. Among these, approximately one quarter are produced in the $\nu\bar{\nu}H$ final state from $ZH\,(Z\to\nu\bar{\nu})$ and $WW$-fusion processes. This channel offers a particularly clean environment in which all visible particles originate from the Higgs decay, and therefore provides a convenient benchmark for the simulation study used in this work.
With the combination of high-performance detectors and advanced particle reconstruction algorithms, the \(H\to c\bar{c}\) is expected to be measured to a percent-level of accuracy, while discovering the \(H\to s\bar{s}\) decay mode is a critical and promising possibility at the future Higgs factory.

The vertex detector is essential for Jet Origin Identification (JOI)~\cite{JOI}, which determines the colored parton from which the jet originates, and thus plays a pivotal role in rare Higgs decay measurements. The characterization of the vertex detector largely depends on its geometric configuration, particularly the innermost radius and spatial resolution. In this work, we quantify the relevant detector metrics such as track impact parameter resolution and JOI performance. Furthermore, we evaluate the precision for $ H\to c\bar{c}$ and significance for $ H\to s\bar{s}$ as functions of these detector parameters, explicitly establishing how detector optimisation translates into physics performance.

This work presents a conceptual detector optimization study in the context of AI-enhanced reconstruction. 
Focusing on the vertex detector geometry, we evaluate performance at three levels—the track reconstruction, JOI, and physics benchmark processes—to provide a coherent view of how geometry optimization propagates from low-level reconstruction to physics sensitivity. 

This paper is structured as follows. In section~\ref{sec:detector}, we introduce the CEPC vertex detector concept, the fast simulation frameworks, the Monte Carlo samples employed. Section~\ref{sec:performance} evaluates detector‐level performance: we first quantify how variations in the vertex detector’s inner radius and spatial resolution affect track impact‐parameter resolution (Section~\ref{subsec:impact_para}), then assess their influence on JOI via migration matrix metrics (Section~\ref{subsec:joi_performance}). We translate these detector and algorithmic results into physics impact by studying the statistical precision of $H\to c\bar c$ and the significance of $H\to s\bar s$ (Section~\ref{subsec:events_analysis}). Section~\ref{sec:summary} summarizes our findings, discusses limitations of the current study, and outlines avenues for further optimisation.

\section{CEPC Vertex Detector and Samples}
\label{sec:detector}
\begin{figure}[htb]
    \centering
     \includegraphics[width=0.5\textwidth]{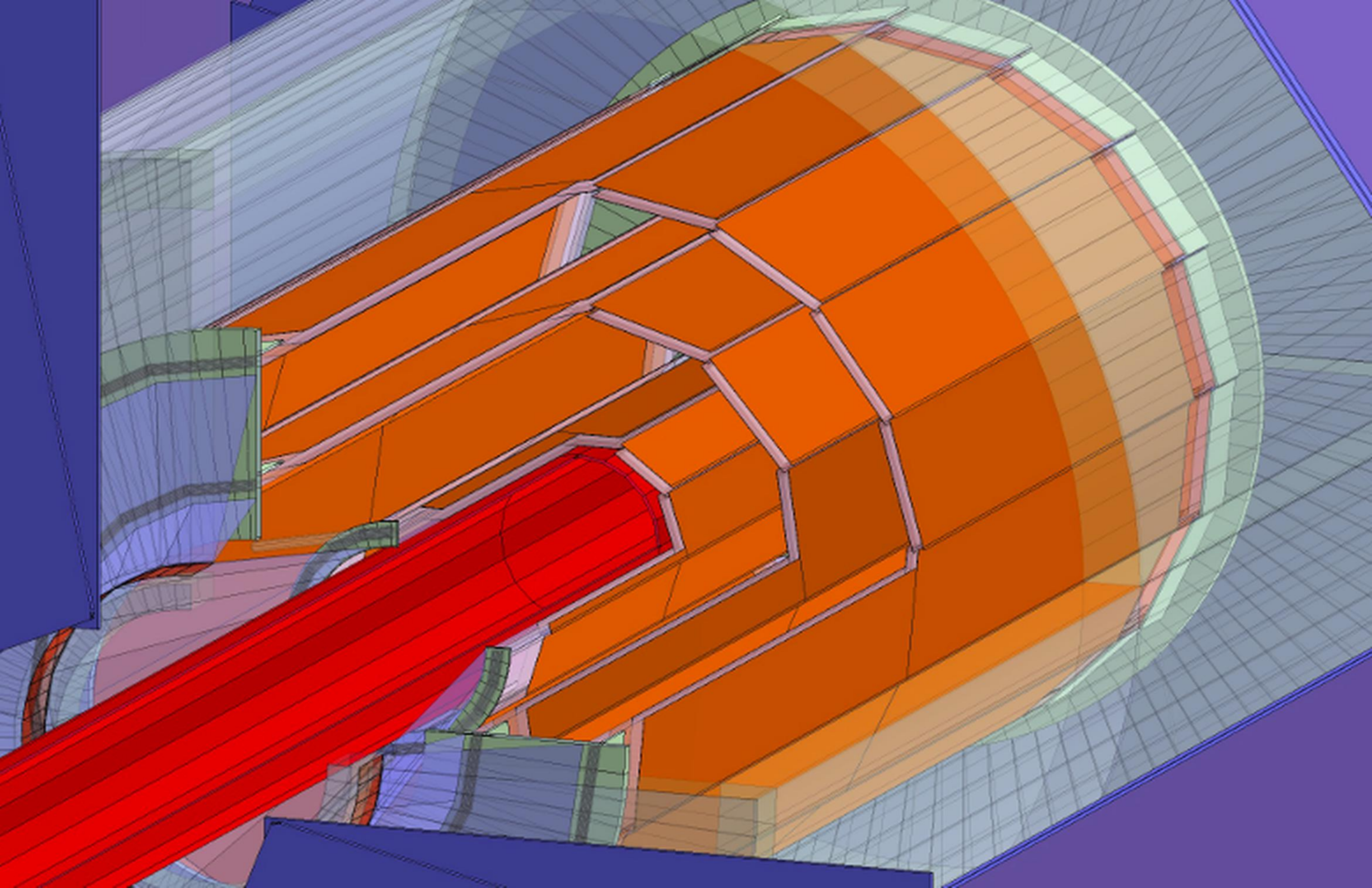}
    \caption{Schematic view of pixel detector. Two layers of silicon pixel sensors are mounted on both sides of each of the three ladders to provide six space points. Only the silicon sensor sensitive region (in orange) is depicted. The vertex detector surrounds the beam pipe (red).}
    \label{fig:vertex_detector}
\end{figure}

\begin{table}[htb]
\centering
\begin{tabular}{lcccc}
\hline
          & $R$ (mm) & $|z|$ (mm) & $|\cos\theta|$ & $\sigma$ ($\mu$m) \\
\hline
Layer 1   & 16       & 62.5       & 0.97           & 2.8                 \\
Layer 2   & 18       & 62.5       & 0.96           & 6                   \\
Layer 3   & 37       & 125.0      & 0.96           & 4                   \\
Layer 4   & 39       & 125.0      & 0.95           & 4                   \\
Layer 5   & 58       & 125.0      & 0.91           & 4                   \\
Layer 6   & 60       & 125.0      & 0.90           & 4                   \\
\hline
\end{tabular}
\caption{Baseline design parameters of the CEPC vertex detector. $R$ denotes the radial position of each layer with respect to the beam line, $|z|$ the half length of the layer, $|\cos\theta|$ the corresponding polar-angle acceptance, and $\sigma$ the spatial resolution of the silicon pixel sensors.}
\label{tab:base_vtx}
\end{table}
The CEPC baseline detector adopts a Particle Flow Oriented design~\cite{Ruan_ArborPerf_2018}, emphasizing the separation and reconstruction of every final state particle, and measures them in the most suited sub-detectors. A crucial component of the CEPC detector is its vertex system.  As referenced in the CEPC Conceptual Design Report (CDR) \cite{CEPC_CDR_Phy}, the vertex system consists of six concentric cylindrical layers of silicon pixel sensors (Figure~\ref{fig:vertex_detector}). The first barrel layer is positioned merely 16~mm from the beam line, and with a per-layer material budget of only \(0.15\%\,X_0\), it attains a spatial resolution of \(2.8\,\mu\text{m}\) for high precision flavour tagging.
The baseline geometrical parameters and spatial resolutions of the vertex detector layers used in this study are summarised in Table~\ref{tab:base_vtx}.
The beam pipe dominates the upstream material of the vertex detector. In the CEPC TDR design~\cite{CEPC2025TDR}, the Machine Detector Interface(MDI) region is explicitly modelled, including a 0.35~mm beryllium beam pipe with a 10~$\mu m$ inner gold coating and a 0.2~mm cooling-water layer, corresponding in total to about 
0.45\%~$X_{0}$. Each vertex layer contributes 
approximately 0.06 -- 0.28\%~$X_{0}$, dominated by the sensors and local supports.  

The beam-induced background is also highly relevant to the design and operation of the vertex detector, as its hits can dominate the occupancy and bandwidth of the detectors. The CEPC TDR provides a dedicated simulation of the beam-induced background, showing that the impact on the impact-parameter resolution is under control. Further details are provided in Appendix~\ref{app:appendix_B}. Hence, the beam-induced background is neglected in this study.

The simulated samples correspond to the $\nu\bar{\nu}H$ final state, including both the $ZH$ process with $Z \to \nu\bar{\nu}$ and the $WW$-fusion contribution ($e^{+}e^{-} \to \nu_e\bar{\nu}_e H$), corresponding to about $0.9\times10^{6}$ Higgs bosons at an integrated luminosity of $20~\mathrm{ab}^{-1}$. We use \textsc{Whizard}~1.95~\cite{whizard} and \textsc{Pythia}~6.4~\cite{pythia6} as the event generators and \textsc{Delphes}~\cite{delphes} for the detector simulation. The reconstructed final-state particles are clustered into two jets via the  $e^+e^-$-$k_t$ algorithm~\cite{Suehara:2015ura, Catani:1991hj}, and JOI is performed to calculate likelihoods of 11 jet categories for jet-flavour tagging. The Delphes-based results are cross-checked against a \textsc{Geant4} based full simulation, demonstrating good agreement and robustness. The details can be found in the appendix \ref{app:appendix_A}. 
An idealised particle identification (PID) model is adopted to study the effects of vertex detector parameters in isolation, without introducing additional dependencies from PID modelling.
Studies of the CEPC baseline detector show that kaon identification efficiencies of about 95\%--98\% can be achieved over a broad kinematic range, approaching the idealised PID assumption~\cite{Zhu2023PID,Yu2025PID}.
This supports the use of idealised PID within the conceptual scope of the present work.

Recent international design studies for future $e^+e^-$ colliders converged on similar requirements for the vertex detector geometry. The ECFA study~\cite{ECFA_DRD} recommends an innermost layer radius down to 16~mm and a spatial resolution of about 3~$\mu$m, with the material budget per layer kept around 0.1\%$X_{0}$. Comparable specifications are adopted in the baseline designs of ILC~\cite{ILC_TDR}, CLIC~\cite{CLIC_CDR}, and FCC-ee~\cite{FCCee_CDR}, targeting optimal flavour tagging and impact parameter resolution.

Motivated by the conclusions of the international studies, we perform parameter scans for the inner radii of the first two vertex layers and the spatial resolution, exploring scaling factors from 0.5 to 2 relative to our baseline detector configuration. This range covers both ambitious future upgrades and more conservative scenarios. The scaling factors are defined as
\begin{equation}
    R_{\text{rad}} \equiv \frac{r_{\text{inner}}}{r_{\text{baseline}}}, \qquad
    R_{\text{res}} \equiv \frac{\sigma_{\text{hit}}}{\sigma_{\text{baseline}}},
\end{equation}
where $r_{\text{inner}}$ and $\sigma_{\text{hit}}$ are the innermost layer radius and spatial resolution. We vary these scaling factors from 0.5 to 2 relative to the baseline detector configuration. Then we assess their impacts at multiple levels:
\begin{itemize}
    \item At the track reconstruction level, we quantify the impact parameter resolution via the root-mean-square (RMS) of its distribution, directly reflecting the detector's ability to localize particle trajectories;
    \item At the JOI performance level, we evaluate JOI using migration matrices for different configurations, and use their traces to characterize overall flavour-tagging efficiency; 
    \item At the higgs benchmark level, we study the efficiency$\times$purity for $H\to c\bar{c}$ and $H\to s\bar{s}$, and quantify the resulting statistical precision for $H\to c\bar{c}$ and signal significance for $H\to s\bar{s}$. 
\end{itemize}

For each Higgs decay channel (\(H\!\to b\bar{b},\,c\bar{c},\,s\bar{s},\,u\bar{u},\,d\bar{d}\) and \(H\!\to gg\)) of $\nu \bar{\nu}H$, we generate one million events to ensure sufficient statistical precision for jet flavour tagging and detector performance studies.

\section{Flavour tagging algorithms and performance under inner radius/spatial resolution scans}
\label{sec:performance}

This section presents a comprehensive evaluation of flavour tagging performance as a function of vertex detector configuration. We first investigate the impact parameter resolution at the track level, followed by the JOI performance under different detector setups. Finally, we connect these detector-level metrics to physics benchmarks, quantifying their influence on $H \to c\bar{c}$ and $H \to s\bar{s}$ analyses.

\subsection{Performance of track reconstruction}
\label{subsec:impact_para}

\begin{figure}[!h]
\centering
\includegraphics[width=0.42\textwidth]
{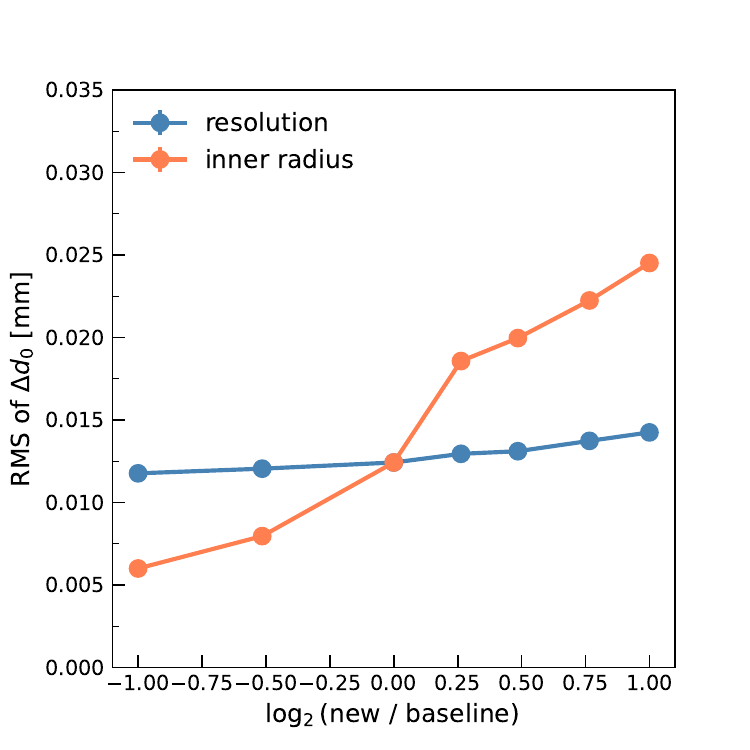}
\includegraphics[width=0.42\textwidth]
{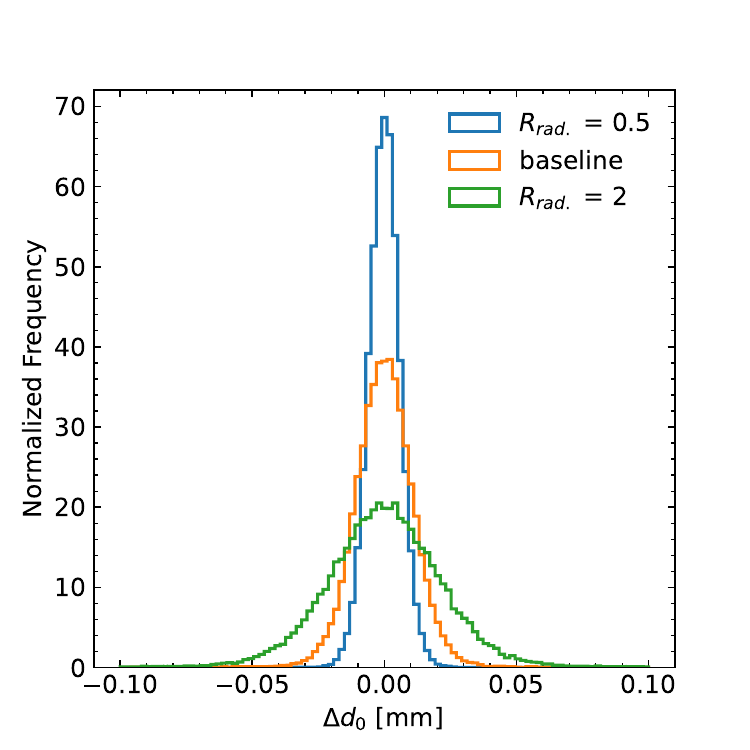}
\includegraphics[width=0.42\textwidth]
{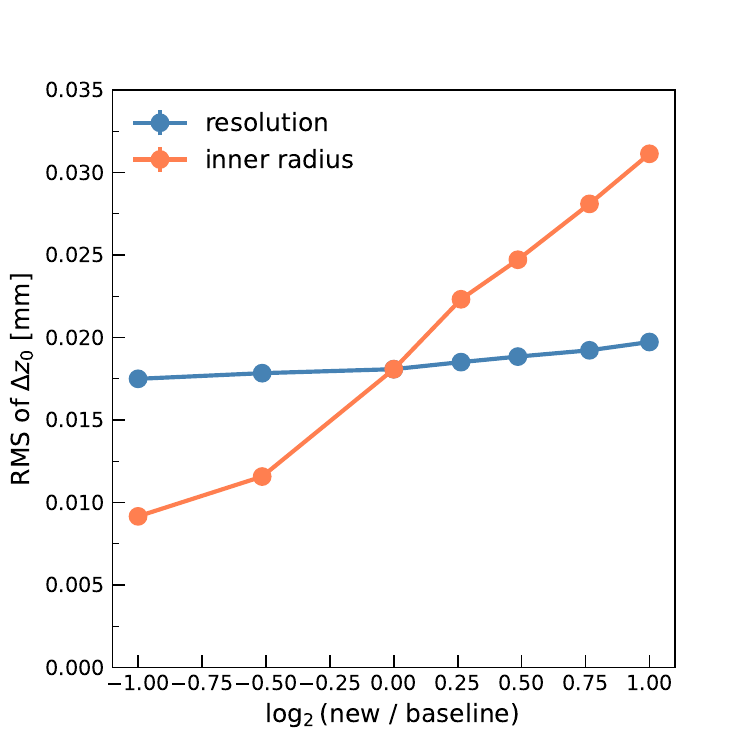}
\includegraphics[width=0.42\textwidth]
{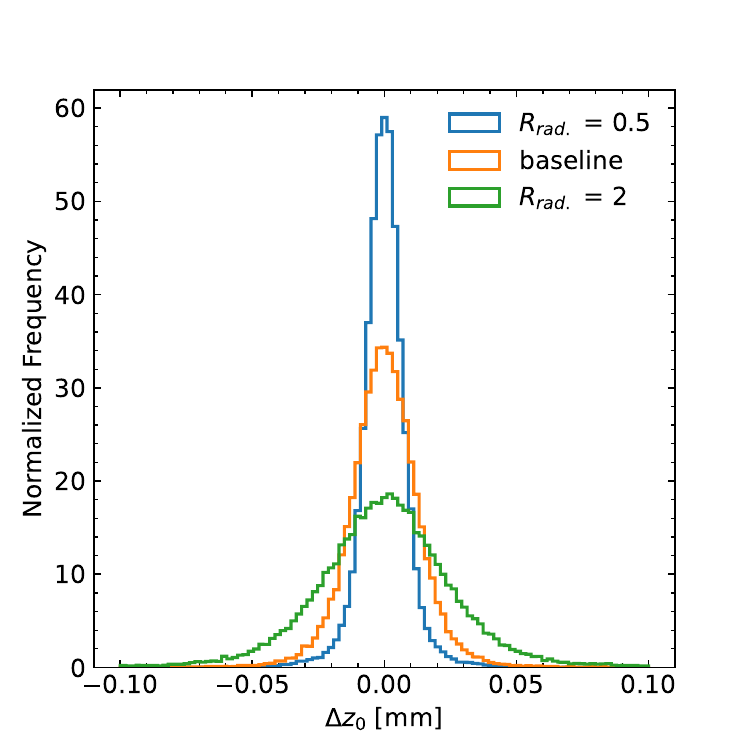}
\caption{Left: RMS of $\Delta d_0$ (top) and $\Delta z_0$ (bottom) versus geometry scale
$\log_2 (\text{new} / \text{baseline}) $.
Right: normalised $\Delta d_0$ (top) and $\Delta z_0$ (bottom) distributions for
$R_{\rm rad}=0.5$, baseline, and $R_{\rm rad}=2$, respectively. The simulated samples are \(H\!\to b\bar{b}\).}
\label{fig:helix_para}
\end{figure}

We quantify the sensitivity of impact parameter resolution to vertex geometry by evaluating the RMS of the $\Delta d_0$ and $\Delta z_0$ distributions, where $\Delta d_0$ and $\Delta z_0$ denote the difference between reconstructed and true transverse and longitudinal impact parameters, respectively, as shown in figure~\ref{fig:helix_para}.

The RMS of $\Delta d_0$ and $\Delta z_0$ depends almost linearly on both the radial radius $R_{\mathrm{rad}}$ and the spatial resolution $R_{\mathrm{res}}$, with $R_{\mathrm{rad}}$ playing the dominant role. Doubling $R_{\mathrm{rad}}$ roughly doubles the impact-parameter resolution, whereas the variations in $R_{\mathrm{res}}$ change it by only about 10\%. As shown in the right panel of figure \ref{fig:helix_para}, reducing $R_{\mathrm{rad}}$ narrows the distributions of $\Delta d_0$ and $\Delta z_0$ (improving the longitudinal resolution by roughly 40\%), while increasing $R_{\mathrm{rad}}$ broadens them.  Thus, the detector’s inner radius is the principal factor in determining impact-parameter performance, and spatial resolution plays only a secondary role in the baseline design.

We also validated our fast simulation results against full \textsc{Geant4}-based simulations. 
A detailed comparison of the impact parameter resolution is provided in appendix~\ref{app:appendix_A}. 
The agreement, as shown in figure~\ref{fig:d0_compare}, demonstrates the robustness of our fast simulation framework.

\subsection{Performance of jet origin identification}
\label{subsec:joi_performance}
The JOI based on the ParticleNet architecture~\cite{Qu:2019gqs} is designed to determine the coloured particle from which a jet originates. For each jet, the algorithm takes the kinematic variables of the reconstructed 
final-state particles, their particle-type information, the impact-parameter measurements of charged tracks, their energy and momentum distributions, as inputs. It then computes likelihood scores for eleven jet-flavour categories (\(b\), \(\bar b\), \(c\), \(\bar c\), \(s\), \(\bar s\), \(u\), \(\bar u\), \(d\), \(\bar d\), and gluon), which are subsequently employed to construct flavour-tagging observables.

Figure~\ref{fig:joi} shows the performance of the JOI algorithm. The left-hand panel displays the normalized 11-dimensional jet-flavour migration matrix ($M_{11}$), obtained by assigning each jet to the flavour category with the highest score, thus giving an overview of the tagging and misidentification pattern among all flavours.
The right-hand panel provides a complementary flavour-specific view by presenting the $c$-jet tagging performance, obtained from scanning working points with different $c$-jet efficiencies. For each chosen $c$-jet efficiency, the corresponding misidentification rates of the other flavour jets are shown. The $c$-tagging efficiency is defined as the fraction of true $c$-jets whose $c$-score exceeds the threshold, while the mis-ID rate is the fraction of non-$c$ jets passing the same threshold. In the efficiency range of $0.2$ to $0.8$, the JOI achieves misidentification rates that are one to three orders of magnitude lower than those of the traditional multivariable analysis approach based on XGBoost~\cite{refTDR_report}.

\settoheight{\matrixheight}{\includegraphics[width=0.45\textwidth]{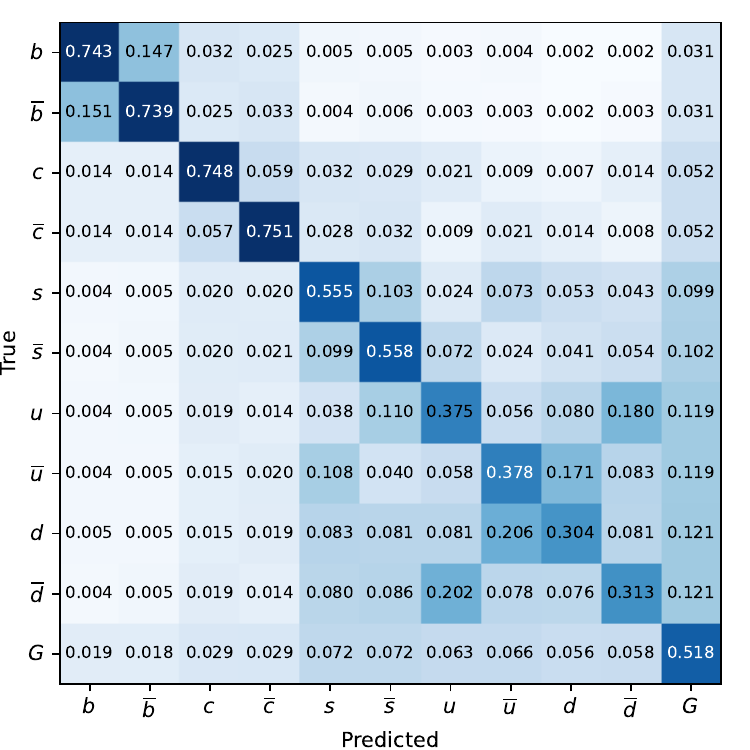}}

\begin{figure}[!htp]
    \centering
    \includegraphics[width=0.45\textwidth]{figures/base_ConfusionMatrix.pdf}
    \includegraphics[height=\matrixheight,width=0.5\textwidth]{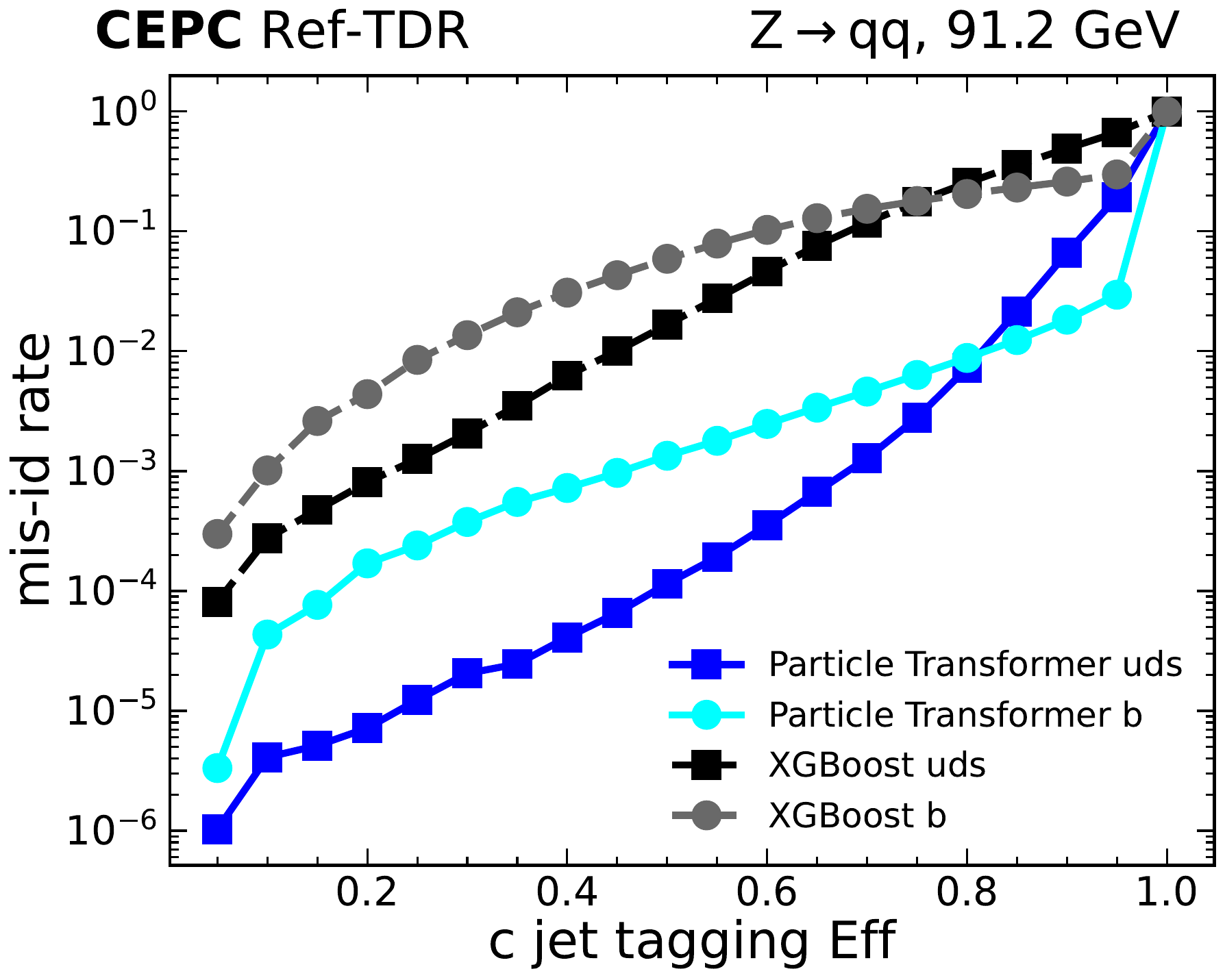}

    \caption{Left: The migration matrix of 11-dimensional jet identification of JOI under the CEPC CDR configuration. The matrix is normalized to unity for each truth label (row). Right: The comparison of JOI and XGboost of the $c$-jet tagging efficiency.~\cite{refTDR_report}}
    \label{fig:joi}
\end{figure}

\begin{figure}[htbp]
  \centering
  \subfloat[$R_{\mathrm{rad}} = 0.5$]{
    \includegraphics[width=0.45\textwidth]{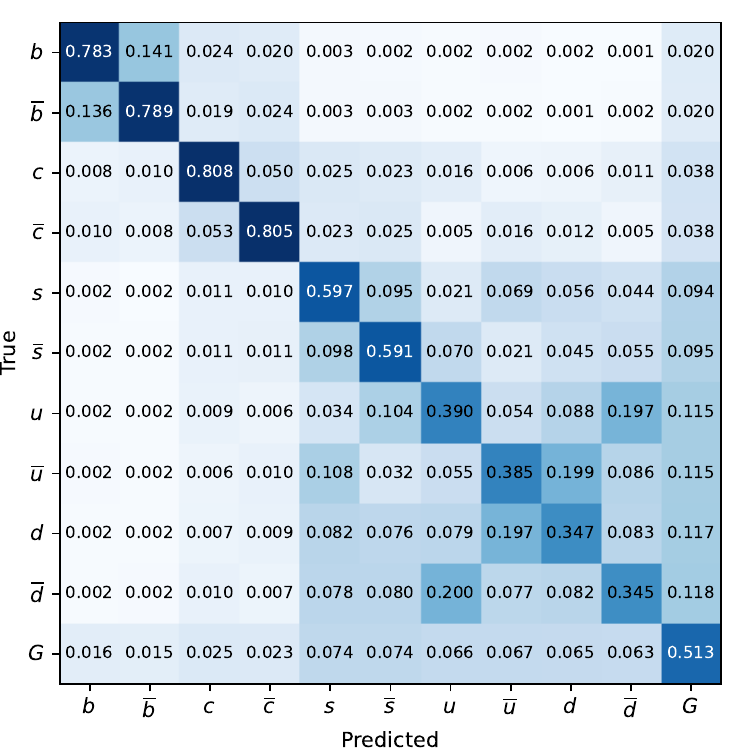}
  }
  \hfill
  \subfloat[$R_{\mathrm{rad}} = 2$]{
    \includegraphics[width=0.45\textwidth]{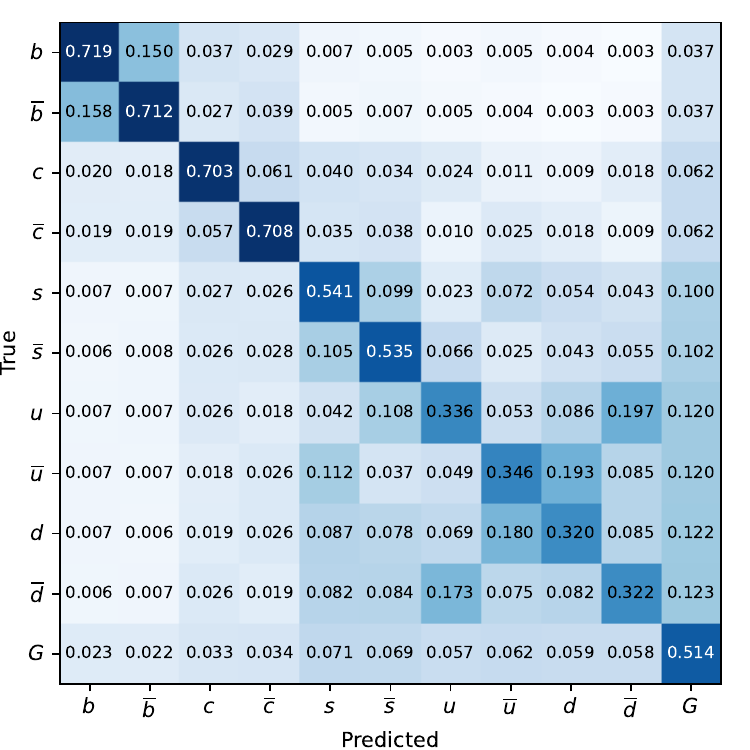}
  }

  \caption{The migration matrix of $M_{11}$ with $R_{rad.}$ = 0.5 (a) and $R_{rad.}$ = 2 (b). The matrix is normalized to unity for each truth label.}
  \label{fig:migration_matrix}
\end{figure}

\begin{figure}[htbp]
\centering
\includegraphics[width=0.45\textwidth,height=\matrixheight]{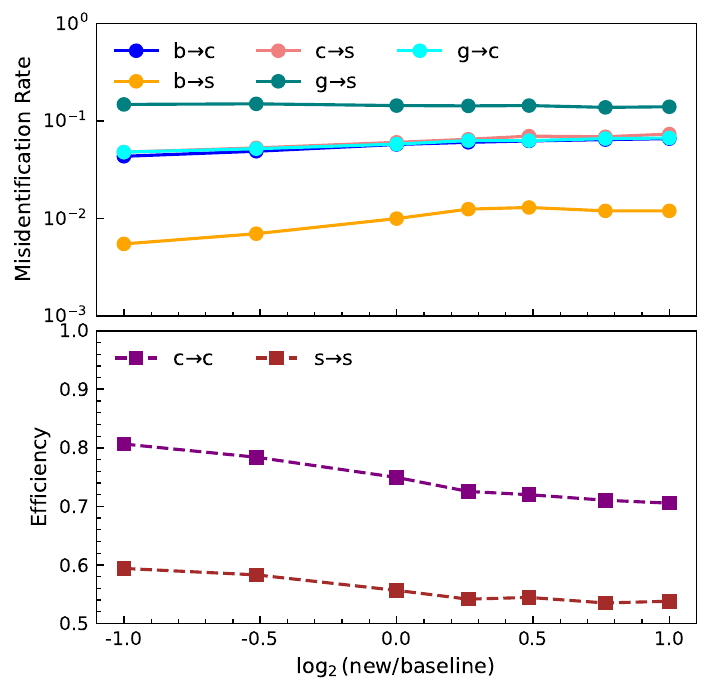}  
\includegraphics[width=0.45\textwidth,height=6.75cm]{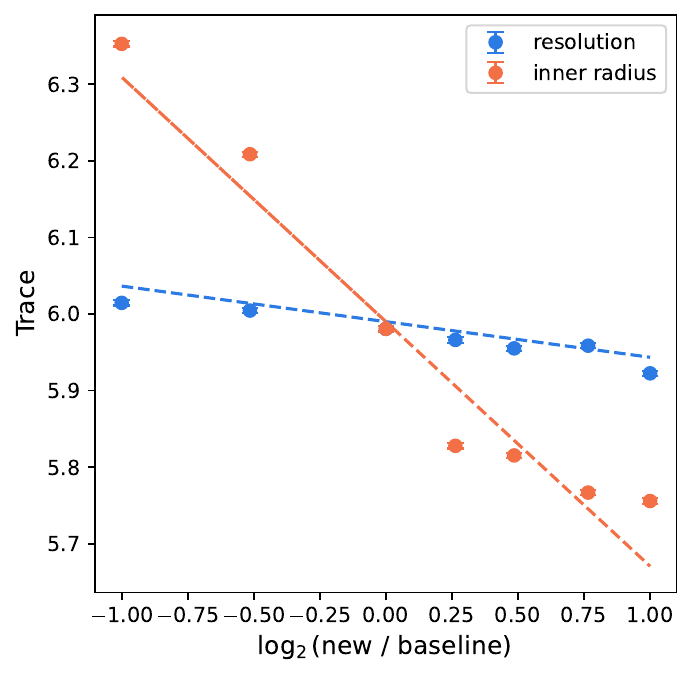}  

\caption{Left: Key misidentification rates and tagging efficiencies versus $\log_2(\text{new}/\text{baseline})$ for different inner radius configurations. 
Right: Trace of the JOI migration matrix versus \(\log_{2}(\text{new}/\text{baseline})\) for different inner radius and resolution configurations.
Orange points indicate scans of the inner radius, while blue points indicate scans of the spatial resolution. Dashed lines represent linear fits to the configurations, respectively. 
}
\label{fig:trace}
\end{figure}

From the migration matrices at varying $R_{\mathrm{rad}}$ as shown in figure~\ref{fig:migration_matrix}, we extract key flavour-tagging metrics as shown on the left-hand side of figure~\ref{fig:trace}. As $R_{\mathrm{rad}}$ increases from $0.5$ to $2$, the $b\to c$, $g\to c$ and $c\to s$  misidentification rate rises from about $4\%$ to $7\%$, Meanwhile, the $c\to c$ tagging efficiency drops from $81\%$ to $70\%$, and $s\to s$ from $60\%$ to $54\%$. While the $g\to s$ rate remains stable around $14\%$. 
The trends indicate that reducing the vertex detector’s innermost radius significantly lowers the $b\to s$ and $c\to s$ misidentification rates, reflecting the crucial role of the vertex in distinguishing heavy-flavour jets from strange jets. In contrast, the impact of vertex detector optimisation on gluon-strange discrimination is limited as gluon jets predominantly originate from the primary vertex, and vertex detector upgrades have little effect on their identification.


To further probe the global performance of JOI, we characterize its behaviour via the trace of \(M_{11}\), denoted as \(\mathrm{Tr}(M)\). This trace, which quantifies the total correct classification probability across all quark flavour categories, is analysed as a function of the geometry scaling factors \(\log_{2} (\text{new} / \text{baseline} )\) for both \(R_{\rm rad}\) and \(R_{\rm res}\). As shown on the right-hand side of figure ~\ref{fig:trace}, a linear regression to the $\mathrm{Tr}(M)$ yields:
\begin{equation}
\mathrm{Tr}(M) = 5.98 \;-\; 0.27\,\log_{2}R_{\rm rad}\;-\; 0.05\,\log_{2}R_{\rm res},    
\end{equation}
demonstrating that variations in the inner radius impact JOI performance by a factor of five more strongly than equivalent changes in spatial resolution.

\subsection{Performance of higgs benchmarks: $\nu \bar{\nu} H$ channel, \(H\to c\bar c\) and \(H\to s\bar s\) measurements }
\label{subsec:events_analysis}

The JOI algorithm provides, for each jet \(j\), a set of posterior probabilities \(P_{\mathrm{jet}}(f\,|\,j)\) that the jet originated from parton flavour \(f\).
To translate these per-jet probabilities into event-level information for $ H\to jj$ decays, we construct a Bayesian likelihood of the Higgs decay hypothesis
\(H\!\to f\bar f\),
\begin{equation}
  L_{f\bar f}(j_{1},j_{2}) \;=\;
  \eta_{f\bar f}\,{\cal B}(H\!\to f\bar f)\,
  \bigl[
        P_{\mathrm{jet}}(f\,|\,j_{1})P_{\mathrm{jet}}(\bar f\,|\,j_{2})
      + P_{\mathrm{jet}}(f\,|\,j_{2})P_{\mathrm{jet}}(\bar f\,|\,j_{1})
  \bigr],
  \label{eq:naiveBayes}
\end{equation}
where \(\eta_{f\bar{f}} = \frac{1}{2}\).  
For quark final states, the factor of \(\frac{1}{2}\) arises because the correction to the branching ratio — where the first jet is \(f\) and the second jet is $\bar{f}$ — accounts for only half of \({\cal B}(f\bar{f})\).  
This factor of \(\frac{1}{2}\) also appears for \(gg\) final states, cancelling out the over-counting due to identical final states.
Normalising over all decay hypotheses gives the event–level posterior
\begin{equation}
  S_{f\bar f}(j_{1},j_{2})
  \;=\;
  \frac{L_{f\bar f}(j_{1},j_{2})}
       {\displaystyle\sum_{f'} L_{f'\bar f'}(j_{1},j_{2})}\;,
  \label{eq:posterior}
\end{equation}
with \(S_{f\bar f}\in[0,1]\) and
\(\sum_{f} S_{f\bar f} = 1\). The corresponding $S_{ff}$ distributions at event-level are shown in Appendix~\ref{app:appendix_C}.
For a given physics measurement, the candidate \(H\to f\bar f\) signal events are defined as those satisfying \(S_{f\bar f}\) over the threshold, where the threshold score is scanned to optimise the product
\(\epsilon \times P\), to balance the signal
retention and background suppression, and thus the optimal measurement sensitivity.
The \emph{efficiency} \(\epsilon\) is defined as the fraction of true \(H\to f\bar f\) signal events whose \(S_{f\bar f}\) value lies above the chosen threshold relative to the total number of simulated signal events. The \emph{purity} \(P\) is the fraction of selected events that truly originate from the \(H\to f\bar f\) decay, i.e.\ the ratio of correctly identified signal events to all selected events including background contributions.

\begin{figure}[htbp]
  \centering

    \includegraphics[width=0.32\textwidth,height=5cm]{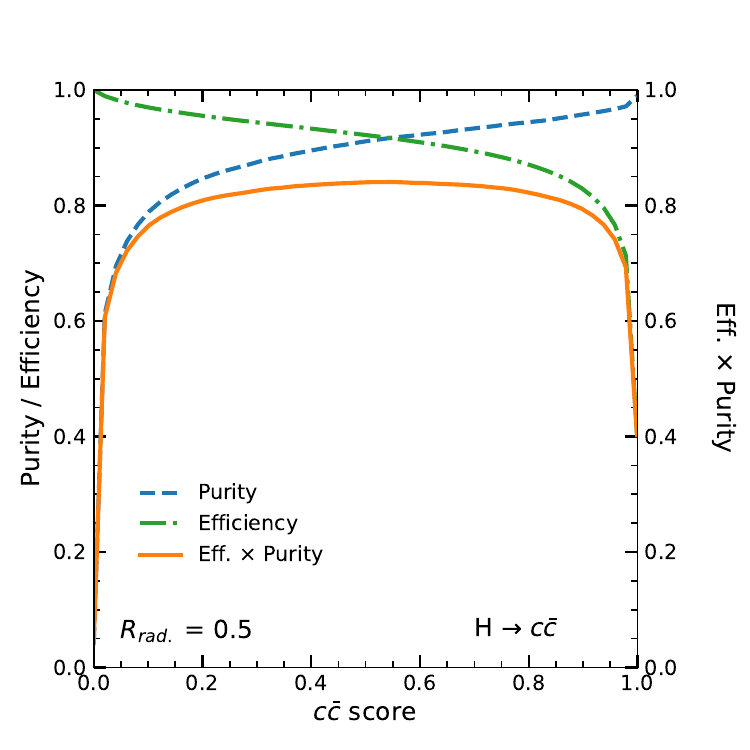}
    \includegraphics[width=0.32\textwidth,height=5cm]{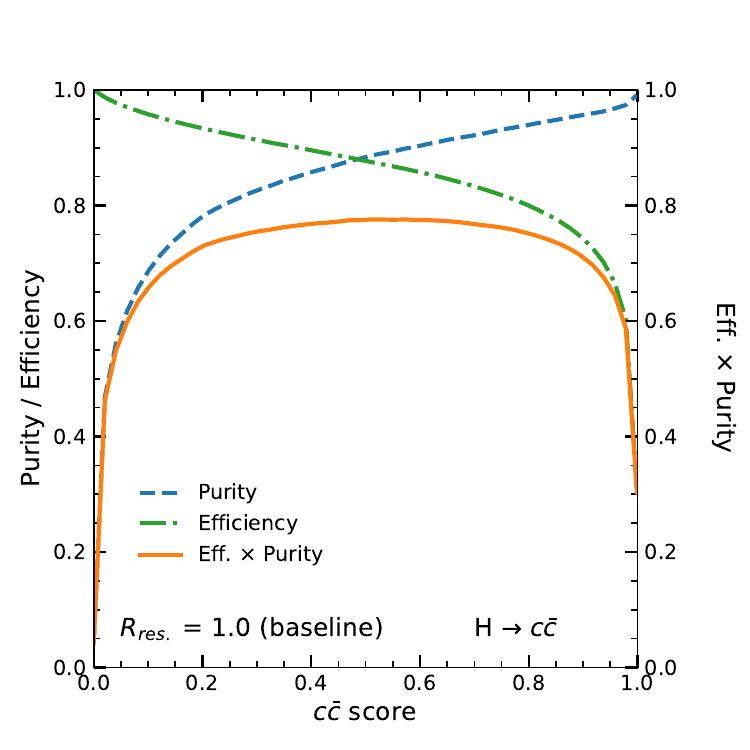}
    \includegraphics[width=0.32\textwidth,height=5cm]{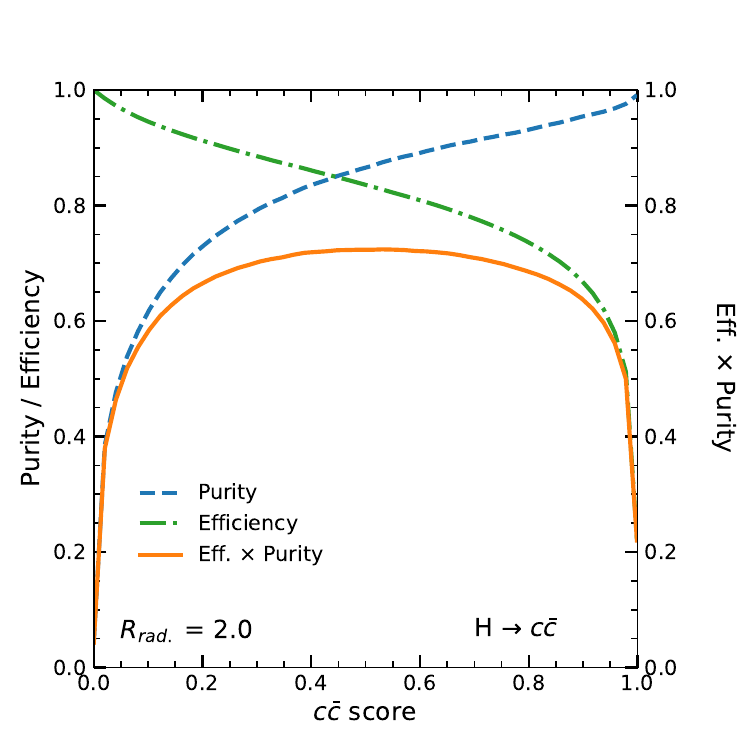}
    \vspace{1em}
    
    \includegraphics[width=0.32\textwidth,height=5cm]{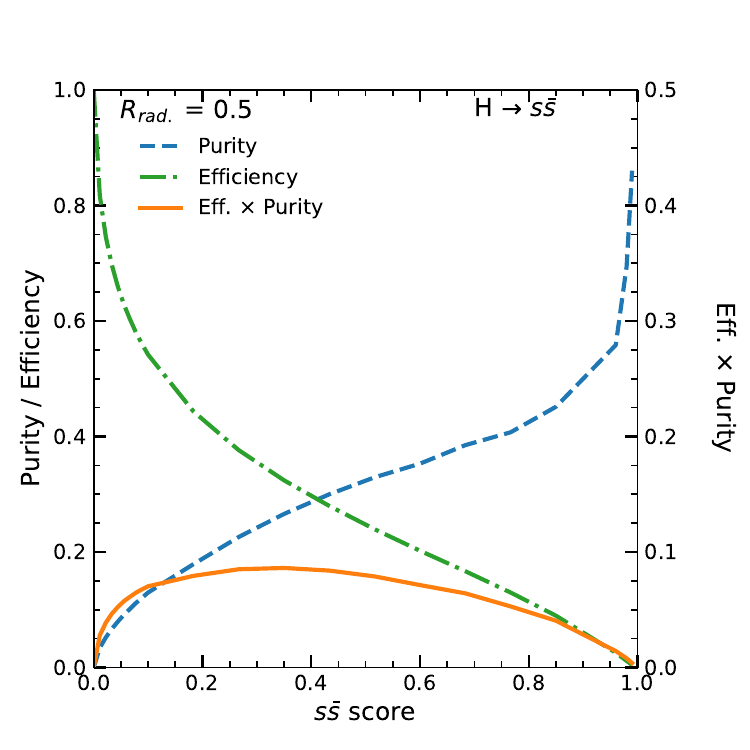}
    \includegraphics[width=0.32\textwidth,height=5cm]{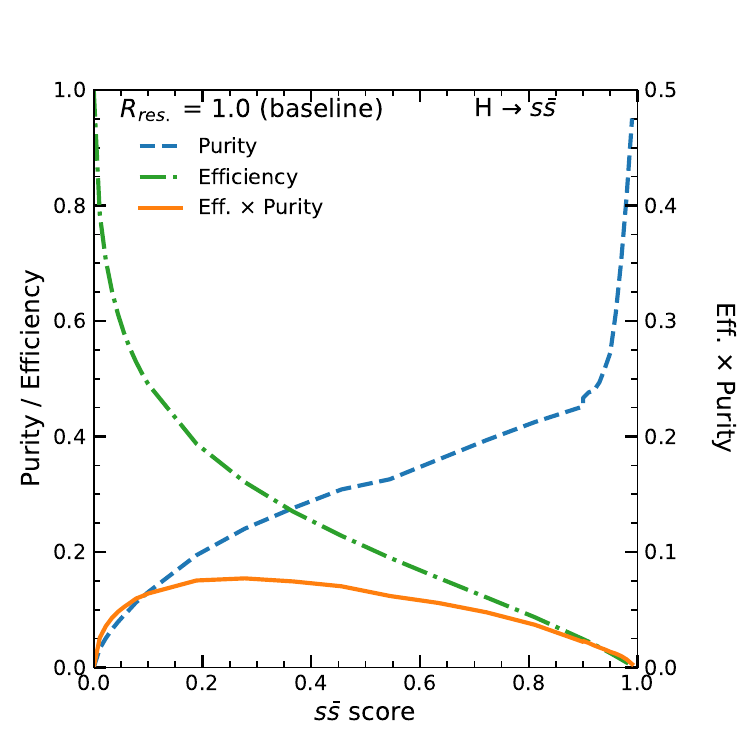}
    \includegraphics[width=0.32\textwidth,height=5cm]{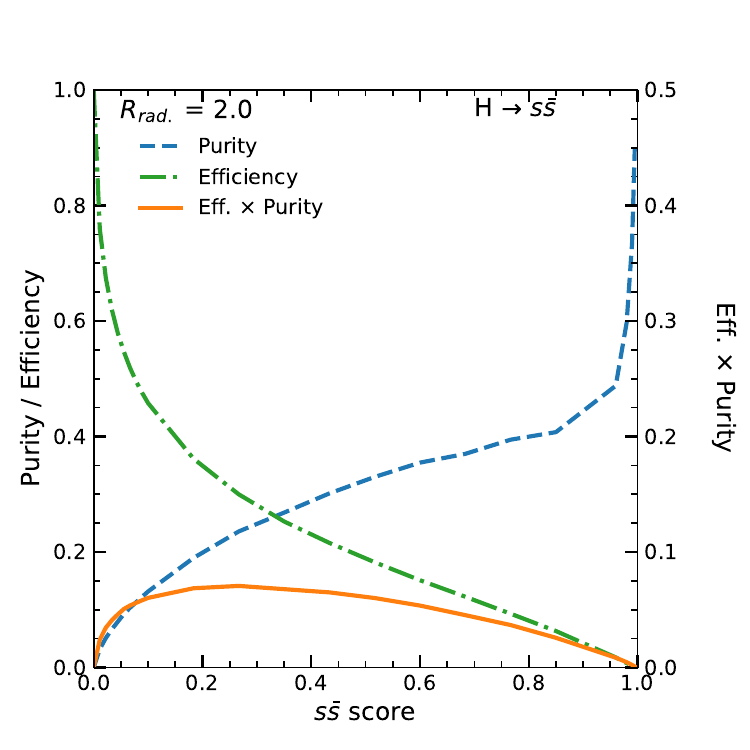}

  \caption{
Purity $P$, efficiency $\epsilon$, and $\epsilon\times P$ as functions of the discriminant score $S_{ff}$ for $H\to c\bar c$ (top) and $H\to s\bar s$ (bottom), shown for $R_{\rm rad}=0.5$ (left), baseline (middle), and $R_{\rm rad}=2$ (right). Note that for the $H\to s\bar s$, the vertical axis for $\epsilon\times P$ is shown in the range $[0,\,0.5]$. The curves are obtained by scanning the $S_{ff}$ selection threshold.}

  \label{fig:EP}
\end{figure}

\begin{figure}[htbp]
  \centering
  
  \begin{minipage}{0.45\linewidth}
    \centering
    \includegraphics[width=\linewidth]{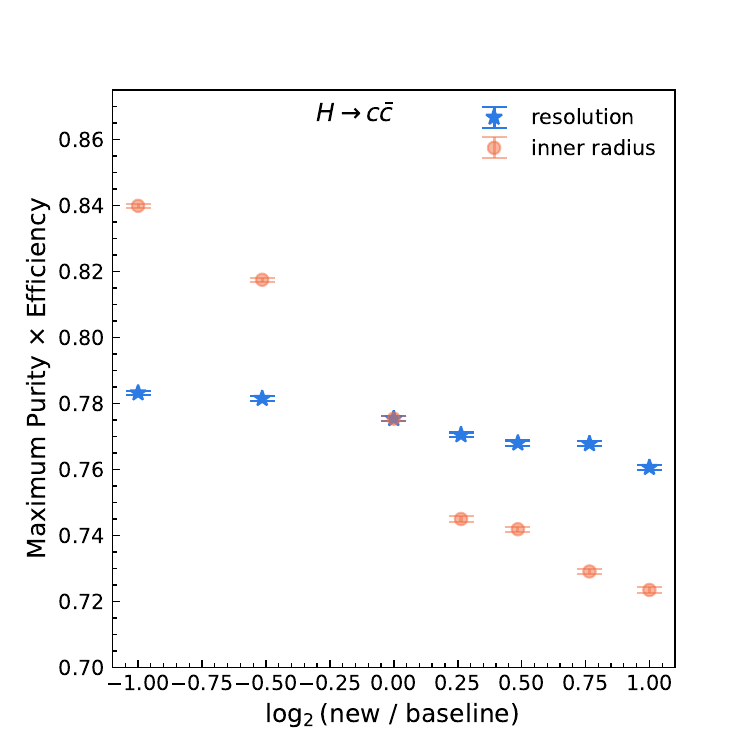}
  \end{minipage}\hfill
  \vspace{1em}
  \begin{minipage}{0.45\linewidth}
    \centering
    \includegraphics[width=\linewidth]{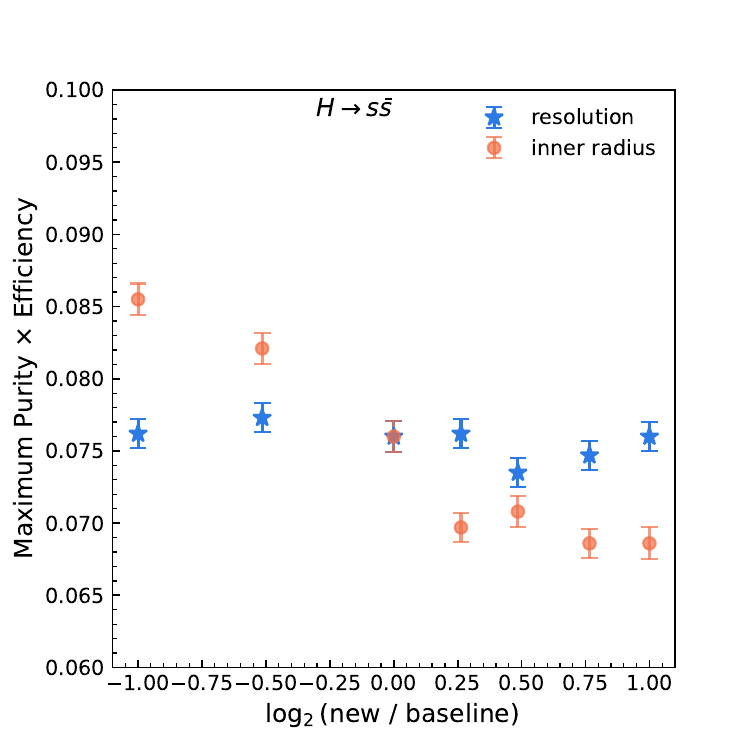}
  \end{minipage}
    \caption{Maximum of \(\varepsilon \times P\) for \(H\to c\bar{c}\) (left) and \(H\to s\bar{s}\) (right). Error bars represent statistical uncertainty.}
  \label{fig:maxEP}
\end{figure}


Using the event-level posterior defined in Eq.~(\ref{eq:posterior}), the purity $P$, efficiency $\varepsilon$, and their product $\varepsilon\times P$ are evaluated as functions of the discriminant score $S_{ff}$. Since previous studies indicate that the impact of the spatial resolution on the flavour-tagging performance is small, figure~\ref{fig:EP} shows results only for three representative values of the vertex inner radius.
For $H\to c\bar c$, the large branching ratio yields a high signal rate and good background separation over a broad range of $S_{cc}$. Both the efficiency and purity remain high for moderately tight selections, resulting in a broad maximum of $\varepsilon\times P$.
In contrast, for $H\to s\bar s$, the maximum of $\varepsilon\times P$ is reached at relatively loose selections. Although the purity improves with increasing $S_{ss}$, the rapid loss of signal efficiency dominates the behaviour of $\varepsilon\times P$, making the optimal working point primarily driven by signal efficiency.



The maximum of \(\varepsilon \times P\) quantifies analysis quality, which is summarized in figure~\ref{fig:maxEP}. In \(H \to c\bar{c}\), the maximum of \(\varepsilon \times P\) scales linearly with \(R_{\text{rad}}\), shifting by 8\% per factor-of-two radius change; \(H \to s\bar{s}\) exhibits similar trends but with larger statistical fluctuations (due to lower signal yield), with the maximum of \(\varepsilon \times P\) varying by ~12\% under the same radius scans. These results reinforce the innermost vertex radius as the primary key point for optimising both decay channels, with pixel resolution showing a marginal impact.

We compute the relative statistical uncertainty for \(H\to c\bar{c}\) and signal significance for \(H\to s\bar{s}\) using the asymptotic approximation for Asimov data. The relative statistical uncertainty for \(H\to c\bar c\) is estimated as eq.(\ref{eq:uncert}).

\begin{equation}
\label{eq:uncert}
      \frac{\delta\mathcal{B}}{\mathcal{B}}
  \;=\;\frac{\sqrt{s+b}}{\,s}\,.
\end{equation}

where $s$ and $b$ denote signal and background yields (extracted from score-based selections).
The significance follows eq.(\ref{eq:sigma})~\cite{Cowan}.
\begin{equation}
\label{eq:sigma}
S = \sqrt{2 \left((s + b) \ln \left(1 + \frac{s}{b}\right) - s\right)}
\end{equation}
where $S$ is the significance.

We evaluate \( H \to c\bar{c} \) yields and statistical uncertainty at the optimal \(\varepsilon \times P\) cut (Table~\ref{tab:sigma_cc}). Halving the inner radius (\( R_{\text{rad}} \)) boosts \( c\bar{c} \) signal, suppresses \( b\bar{b} \) background, then lowers uncertainty by approximately 4\%, while spatial resolution (\( R_{\text{res}} \)) has minimal impact on precision. For \( H \to s\bar{s} \) summarised in Table~\ref{tab:sigma_ss}, the JOI combined with normalized score cuts fully suppresses \( b\bar{b} \) background (\( b\bar{b} \) yields to 0), providing a critical advantage for signal isolation. Decreasing \( R_{\text{rad}} \) suppresses \( gg \) background and enhances significance by 8\% compared to the baseline. Varying \( R_{\text{res}} \) causes negligible changes in signal/background yields and significance, indicating its marginal impact. These trends illustrate that the inner vertex radius dominates sensitivity for both decays, with resolution playing a minor role. 

\begin{table}[h]
  \centering
  \begin{tabular}{lccccc}
    \hline
    Configuration & \(c\bar{c}\,(10^{3})\) & \(b\bar{b}\,(10^{3})\) &
           \(s\bar{s}\) & \(gg\,(10^{3})\) &
           Uncertainty\((10^{-3})\) \\
    \hline
    baseline              & 24.4 & 1.26 & 11.0 & 1.92 & 6.81 \\
    \(R_{\rm res}=0.5\)   & 24.5 & 1.21 &  8.95 & 1.85 & 6.78 \\
    \(R_{\rm res}=2\)     & 24.0 & 1.21 & 12.9 & 1.99 & 6.88 \\
    \(R_{\rm rad}=0.5\)   & 25.6 & 0.75 &  5.54 & 1.75 & 6.55 \\
    \(R_{\rm rad}=2\)     & 23.2 & 1.54 &  1.61 & 2.04 & 7.05 \\
        \hline
  \end{tabular}
    \caption{Event yields at the optimal cut (maximizing $\varepsilon \times P$) and statistical uncertainty for $H \to c\bar{c}$, normalized to an integrated luminosity of $20\,\mathrm{ab}^{-1}$ in the $\nu\bar{\nu}H$ channel, corresponding to about $0.9\times10^{6}$ Higgs bosons. Columns list the signal ($c\bar{c}$), other Higgs decay backgrounds ($b\bar{b}$, $s\bar{s}$, $gg$), and the relative statistical uncertainty.}
  \label{tab:sigma_cc}
\end{table}

\begin{table}[h]
  \centering
  \begin{tabular}{lccccc}
    \hline
    Configuration & \(s\bar{s}\) & \(b\bar{b}\) & \(c\bar{c}\) & \(gg\) & Significance \\
    \hline
    baseline              & 70.9 & 0 & 10.0 & 224.6 & 4.42\,$\sigma$ \\
    \(R_{\rm res}=0.5\)   & 64.1 & 0 &  7.6 & 179.4 & 4.45\,$\sigma$ \\
    \(R_{\rm res}=2\)     & 64.4 & 0 & 10.3 & 181.4 & 4.41\,$\sigma$ \\
    \(R_{\rm rad}=0.5\)   & 70.2 & 0 &  6.2 & 189.4 & 4.76\,$\sigma$ \\
    \(R_{\rm rad}=2\)     & 64.8 & 0 & 11.5 & 203.2 & 4.23\,$\sigma$ \\
    \hline
  \end{tabular}
    \caption{Event yields at the optimal cut (maximizing \(\varepsilon \times P\)) and corresponding Asimov significance for \(H \to s\bar{s}\), normalized to an integrated luminosity of \(20\,\text{ab}^{-1}\) in the $\nu\bar{\nu}H$ channel, corresponding to about $0.9\times10^{6}$ Higgs bosons. Columns list signal (\(s\bar{s}\)), other Higgs decay backgrounds (\(b\bar{b}\), \(c\bar{c}\), \(gg\)), and significance.}
  \label{tab:sigma_ss}
\end{table}


\newpage
\section{Summary and Discussion}
\label{sec:summary}

The vertex detector is crucial for high-precision Higgs measurements at future electron-positron Higgs factories, particularly impacting the sensitivity to rare decays such as \(H \rightarrow c\bar{c}\) and \(H \rightarrow s\bar{s}\).
Precise measurements of these rare decays provide a decisive test of the Yukawa coupling predicted by the Standard Model.

This study, obtained using the AI-driven JOI framework which leverages jet momentum and impact parameters for effective signal discrimination~\cite{JOI}, highlights the inner radius as the primary geometric parameter influencing the impact parameter resolution and subsequently flavour-tagging performance, while spatial resolution exerts a comparatively weaker influence.
Benchmarking the \(\nu\bar{\nu}H\) production channel, we project a statistical precision of 0.66–0.71\% for \(H \rightarrow c\bar{c}\), and a significance exceeding 4$\sigma$ for \(H \rightarrow s\bar{s}\). The measurement sensitivity is moderately affected by the vertex detector design: reducing the inner radius from the baseline leads to less than 10\% improvements in both precision and significance, while increasing it has the opposite effect. In contrast, variations in spatial resolution have a much smaller impact, with only 1\% changes observed across all benchmarks.

Our study is based on fast simulations and currently considers only the major backgrounds from other Higgs decay modes, representing an optimistic yet well-motivated setup. Comparisons with full (\textsc{Geant4}-based) simulations 
confirm the reliability of our fast-simulation approach, with discrepancies at the 5\% level. In a realistic environment, beam-induced and other Standard Model 
backgrounds would further dilute the sensitivity, and their quantitative impact, particularly the contributions related to the MDI design, still requires dedicated study. 
Consequently, the present results should be interpreted as upper-limit estimates achievable under idealised background conditions.


Our observations are consistent with previous studies~\cite{wu_vertex, Zhu2023ParticleNet}, which show that the inner radius of the vertex detector is the most sensitive geometric parameter for flavour tagging and Higgs precision measurements, underscoring its importance in future detector design.
The search for $H \to s\bar{s}$ offers a unique probe of second-generation Yukawa couplings and is a major objective for next-generation Higgs factories. Recent global studies project measurement accuracies for this channel in the range of 17\% to 140\% \cite{Zhu:2025eoe, JOI, ECFA}, with the spread arising from differences in assumed luminosity, analysis strategies, background modelling, and flavour tagging performance.

Our study demonstrates that, within the scanned parameter space, the achievable sensitivity to \(H \rightarrow s\bar{s}\) can vary by more than 12\% due solely to vertex detector design. With an optimized inner radius, advances in detector design and analysis will further enhance the strange-quark Yukawa coupling reach. These improvements include better identification of hadrons containing s-quarks (such as \({K}_{S}^0\) and \(\Lambda\)~\cite{Zheng:2020qyh}), optimizing operation scenarios to increase Higgs boson yields, refining and validating more precise hadronization models, enhancing the performance of flavour-tagging algorithms, leveraging advanced analysis techniques, and incorporating more Higgs production channels (\(l^{+}l^{-}H\), \(q\bar{q}H\)). Collectively, these improvements could push the reach well beyond the current 4$\sigma$ level, potentially enabling its first unambiguous observation and delivering a stringent test of second-generation Yukawa couplings.


\acknowledgments
We thank Yongfeng Zhu for providing the analysis tools and the useful discussions. This work was supported by National Key Program for S$\&$T Research and Development under Contract No. 2024YFA1610603, 2024YFA1610600, National Key R$\&$D Program of China under Contracts No. 2022YFE0116900, the National Key Laboratory of Dark Matter Physics and the National Natural Science Foundation of China under Grant No.W2441004. 

\appendix

\section{Comparison between Delphes and Geant4}
\label{app:appendix_A}

\begin{figure}[htbp]
\centering
    \includegraphics[width=0.45\linewidth]{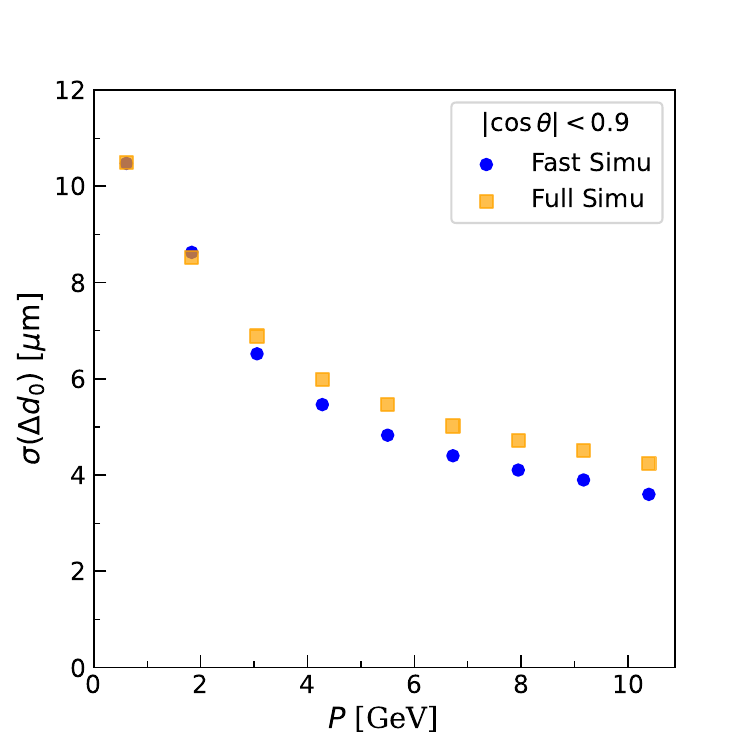}
    \label{fig:sima_d0}
    \caption{Comparison of the transverse impact parameter resolution $\sigma(\Delta d_0)$ as a function of track momentum $P$ in the central region ($|\cos\theta|<0.9$) of $H \to bb$ events between fast simulation(\textsc{Delphes}) and full simulation (\textsc{Geant4}). }
    \label{fig:d0_compare}
\end{figure}

To validate the accuracy of our fast simulation approach, we compare its results with those from the full \textsc{Geant4}-based simulation. Figure~\ref{fig:d0_compare} shows this comparison for the transverse impact parameter ($d_0$) resolution dependence on track momentum ($P$). We observe that the difference between the fast simulation and the full simulation is less than 5\%, confirming that the fast simulation provides a reliable approximation to the full simulation.

\begin{figure}[htbp]
    \centering
    \includegraphics[width=0.45\linewidth]{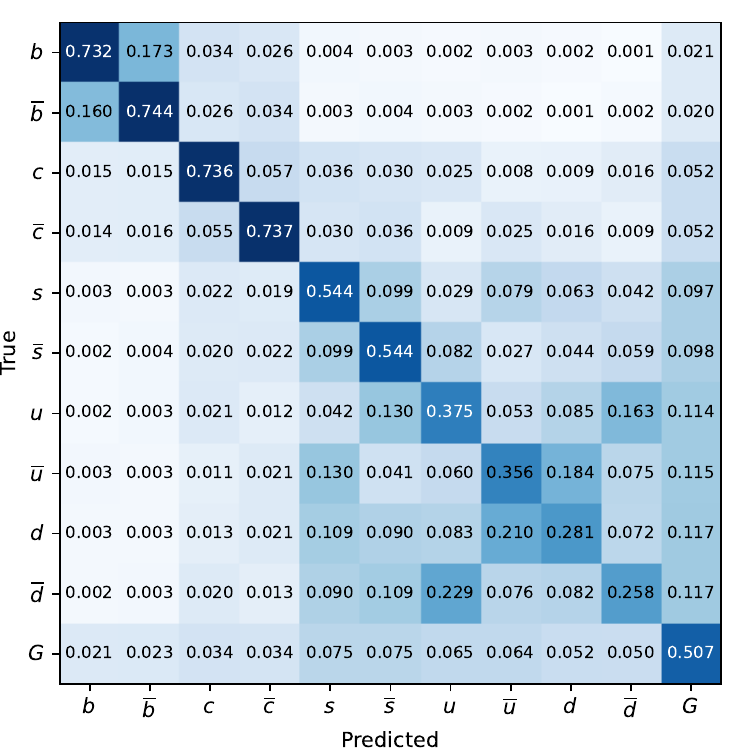}
    \includegraphics[width=0.45\linewidth]{figures/base_ConfusionMatrix.pdf}
    \caption{The migration matrix of $M_{11}$ with full simulation by \textsc{Geant4} (left) and fast simulation \textsc{Delphes} (right).}
    \label{fig:CM_compare}
\end{figure}

Figure~\ref{fig:CM_compare} compares the $M_{11}$ jet–flavour migration matrices obtained with the full \textsc{Geant4} simulation and the fast \textsc{Delphes} simulation.  
For all quark flavours, the diagonal elements differ by less than $3\%$, and the off–diagonal mis-identification rates show no systematic bias. In particular, the difference of charm-jet and strange-jet tagging efficiency is within 2\%.
This close correspondence validates the use of the \textsc{Delphes} framework for the vertex–detector optimisation presented in this work.

\section{Simulation Study of the Beam-Induced Background Impact}
\label{app:appendix_B}

\begin{figure}[htbp]
    \centering
    \includegraphics[width=0.75\textwidth]{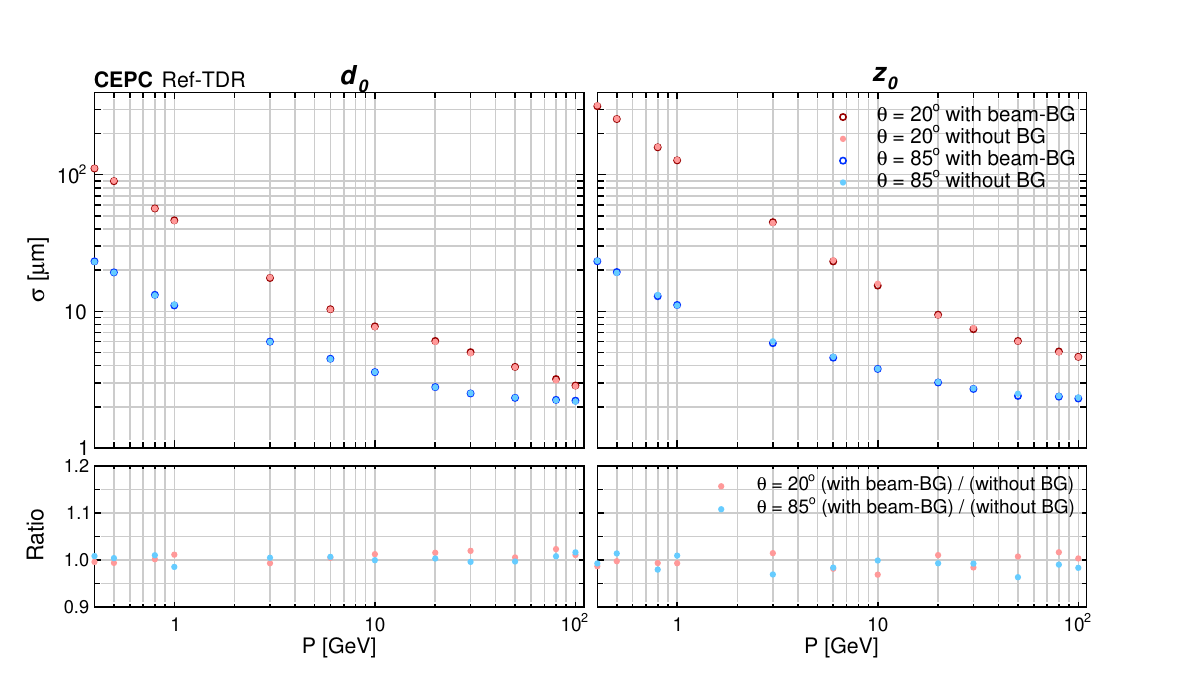}
    \caption{
    Comparison of the impact parameter resolutions ($d_0$, $z_0$) with and without beam-induced background overlay at polar angles of $85^\circ$ and $20^\circ$.}
    \label{fig:bib_d0z0}
\end{figure}

The influence of beam-induced background on tracking performance has been qualitatively evaluated in the CEPC TDR~\cite{CEPC2025TDR}. As shown in Figure~\ref{fig:bib_d0z0}, a direct comparison of the impact-parameter resolutions ($d_0$ and $z_0$) with and without beam-induced background overlay demonstrates no noticeable degradation across the momentum range. Therefore, the pixel size and spatial resolution variations studied in this work remain valid, and the overall conclusions on vertex detector optimization are not affected by beam background effects.

\section{Event-level distributions of the discriminant $S_{ff}$}
\label{app:appendix_C}

\begin{figure}[htbp]
    \centering
    \includegraphics[width=0.40\linewidth]{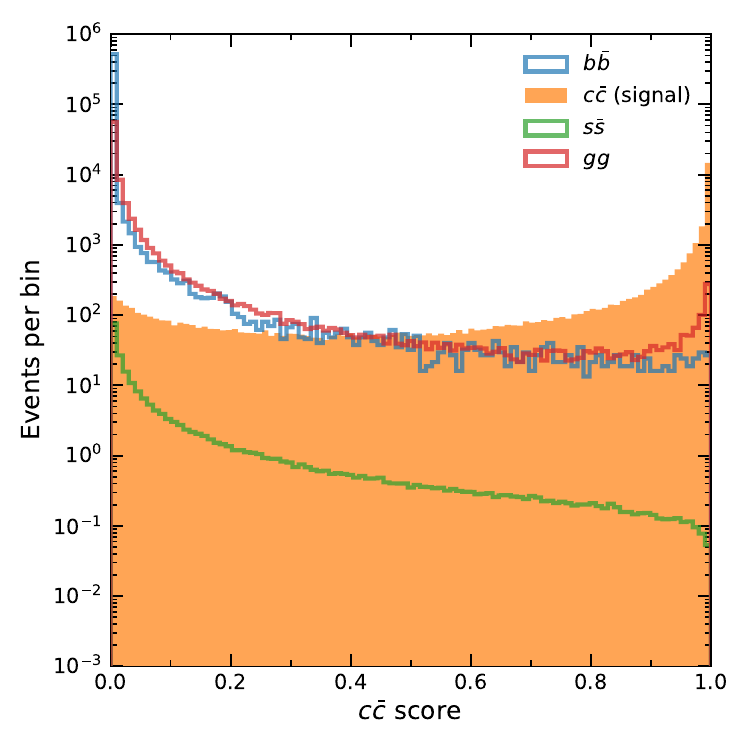}
    \includegraphics[width=0.40\linewidth]{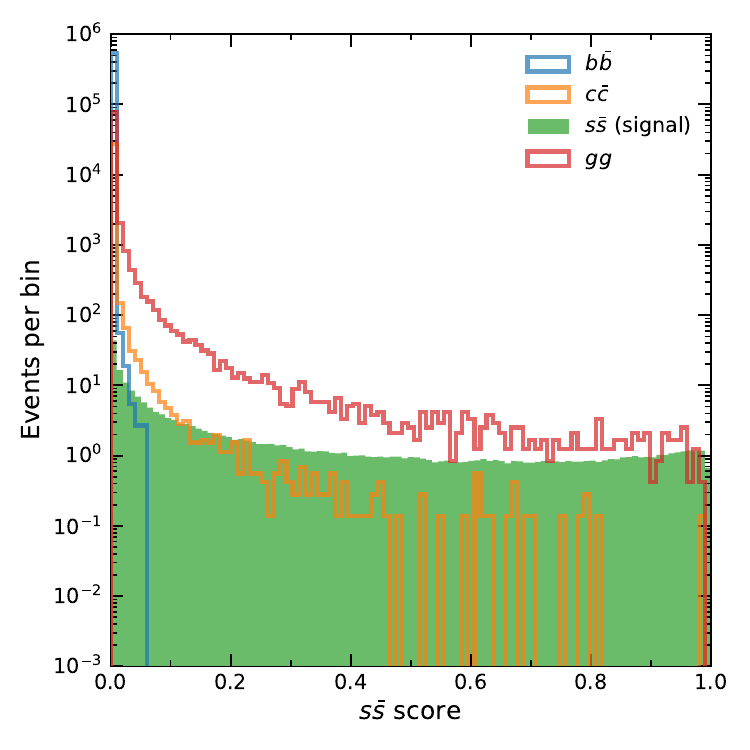}
    \caption{Distributions of the event-level discriminant $S_{ff}$ for the baseline vertex detector configuration.
    Left: $S_{cc}$ distribution for the $H\to c\bar c$.
    Right: $S_{ss}$ distribution for the $H\to s\bar s$.
    The signal and the main Higgs decay backgrounds are shown separately.
    The vertical axis indicates the number of events per bin on a logarithmic scale.}
    \label{fig:sff_dist}
\end{figure}

To provide a more intuitive interpretation of the efficiency--purity curves shown in Figure~\ref{fig:EP}, we present in Figure~\ref{fig:sff_dist} the distributions of the event-level discriminant $S_{ff}$ for the signal and the main Higgs decay backgrounds. The horizontal axis corresponds to the value of $S_{ff}$, while the vertical axis shows the number of events per bin.
As discussed in the main text, varying the selection on $S_{ff}$ corresponds to integrating these distributions above the chosen threshold. Tightening the selection requirement progressively shifts the selected events towards the right-hand side, while looser selections include an increasing fraction of events at small $S_{ff}$ values.

\bibliographystyle{JHEP}
\bibliography{ref.bib}

@techreport{ES,
      author        = {{The European Strategy Group}},
      title         = "{Deliberation document on the 2020 Update of the European
                       Strategy for Particle Physics}",
      reportNumber  = "CERN-ESU-014",
      address       = "Geneva",
      year          = "2020",
      url           = "https://cds.cern.ch/record/2720131",
      doi           = "10.17181/ESU2020Deliberation",
}

@article{SM_BR,
    author = "de Florian, D. and others",
    collaboration = "LHC Higgs Cross Section Working Group",
    title = "{Handbook of LHC Higgs Cross Sections: 4. Deciphering the Nature of the Higgs Sector}",
    eprint = "1610.07922",
    archivePrefix = "arXiv",
    primaryClass = "hep-ph",
    reportNumber = "CERN-2017-002-M, CERN-2017-002",
    doi = "10.23731/CYRM-2017-002",
    journal = "CERN Yellow Rep. Monogr.",
    volume = "2",
    pages = "1--869",
    year = "2017"
}

@article{Hss_BR,
  title = {Probing the Higgs--strange-quark coupling at ${e}^{+}{e}^{\ensuremath{-}}$ colliders using light-jet flavor tagging},
  author = {Duarte-Campderros, J. and Perez, G. and Schlaffer, M. and Soffer, A.},
  journal = {Phys. Rev. D},
  volume = {101},
  issue = {11},
  pages = {115005},
  numpages = {6},
  year = {2020},
  month = {Jun},
  publisher = {American Physical Society},
  doi = {10.1103/PhysRevD.101.115005},
  url = {https://link.aps.org/doi/10.1103/PhysRevD.101.115005}
}

@inproceedings{snowmass,
    author = "Cheng, Huajie and others",
    collaboration = "CEPC Physics Study Group",
    title = "The Physics potential of the CEPC. Prepared for the US Snowmass Community Planning Exercise (Snowmass 2021)",
    booktitle = "{Snowmass 2021}",
    eprint = "2205.08553",
    archivePrefix = "arXiv",
    primaryClass = "hep-ph",
    month = "5",
    year = "2022"
}

@article{CEPC_CDR_Acc,
    author = "CEPC Study Group",
    title={CEPC Conceptual Design Report: Volume 1 - Accelerator}, 
    author={The CEPC Study Group},
    year={2018},
    eprint={1809.00285},
    archivePrefix={arXiv},
    primaryClass={physics.acc-ph},
    url={https://arxiv.org/abs/1809.00285}, 
}

@article{CEPC_CDR_Phy,
    author = "CEPC Study Group",
    editor = "Guimar\~aes da Costa, Jo\~ao Barreiro and others",
    collaboration = "CEPC Study Group",
    title = "{CEPC Conceptual Design Report: Volume 2 - Physics \& Detector}",
    eprint = "1811.10545",
    archivePrefix = "arXiv",
    primaryClass = "hep-ex",
    reportNumber = "IHEP-CEPC-DR-2018-02, IHEP-EP-2018-01, IHEP-TH-2018-01",
    month = "11",
    year = "2018"
}

@article{JOI,
    title       = "{Jet-Origin Identification and Its Application at an Electron-Positron Higgs Factory}",
    author      = "Liang, Hao and Zhu, Yujie and Wang, Yifan and Che, Yuzhi and Ruan, Manqi and Zhou, Cong and Qu, Huilin",
    journal     = "Phys. Rev. Lett.",
    volume      = "132",
    pages       = "221802",
    year        = "2024",
    doi         = "10.1103/PhysRevLett.132.221802",
    url         = "https://doi.org/10.1103/PhysRevLett.132.221802"
}

@article{WHIZARD,
  title = {WHIZARD—simulating multi-particle processes at {LHC} and {ILC}},
  author = {Kilian, Wolfgang and Ohl, Thorsten and Reuter, Jürgen},
  journal = {The European Physical Journal C},
  volume = {71},
  number = {1},
  pages = {1742},
  year = {2011},
  doi = {10.1140/epjc/s10052-011-1742-y}
}

@article{pythia6,
   title={PYTHIA 6.4 physics and manual},
   volume={2006},
   ISSN={1029-8479},
   url={http://dx.doi.org/10.1088/1126-6708/2006/05/026},
   DOI={10.1088/1126-6708/2006/05/026},
   number={05},
   journal={Journal of High Energy Physics},
   publisher={Springer Science and Business Media LLC},
   author={Sjöstrand, Torbjörn and Mrenna, Stephen and Skands, Peter},
   year={2006},
   month=may, pages={026–026} 
}

@article{Cowan,
    title       = "{Asymptotic formulae for likelihood-based tests of new physics}",
    author      = "Cowan, Glen and Cranmer, Kyle and Gross, Eilam and Vitells, Ofer",
    journal     = "Eur. Phys. J. C",
    volume      = "71",
    pages       = "1554",
    year        = "2011",
    doi         = "10.1140/epjc/s10052-011-1554-0",
    url         = "https://doi.org/10.1140/epjc/s10052-011-1554-0"
}

@article{Zhu2023ParticleNet,
  title         = "ParticleNet and its application on CEPC Jet Flavor Tagging",
  author        = {Yongfeng Zhu and Hao Liang and Yuexin Wang and Huilin Qu and Chen Zhou and Manqi Ruan},
  journal       = {arXiv preprint},
  eprint        = {2309.13231},
  archivePrefix = {arXiv},
  primaryClass  = {hep-ex},
  doi           = {10.48550/arXiv.2309.13231},
  url           = {https://arxiv.org/abs/2309.13231},
  year          = {2023}
}

@article{wu_vertex,
author = {Wu, Z. and Li, G. and Yu, Dachaun and Fu, C. and Ouyang, Qiuzi and Ruan, Manqi},
year = {2018},
month = {09},
pages = {T09002-T09002},
title = {Study of vertex optimization at the CEPC},
volume = {13},
journal = {Journal of Instrumentation},
doi = {10.1088/1748-0221/13/09/T09002}
}

@article{an2019precision,
    author = "An, Fenfen and others",
    title = "{Precision Higgs physics at the CEPC}",
    eprint = "1810.09037",
    archivePrefix = "arXiv",
    primaryClass = "hep-ex",
    reportNumber = "FERMILAB-PUB-18-573-T",
    doi = "10.1088/1674-1137/43/4/043002",
    journal = "Chin. Phys. C",
    volume = "43",
    number = "4",
    pages = "043002",
    year = "2019"
}

@article{Zheng:2020qyh,
    author = "Zheng, Taifan and Wang, Jike and Shen, Yuqiao and Cheung, Yeuk-Kwan E. and Ruan, Manqi",
    title = "{Reconstructing $K^0_S$ and $\Lambda $ in the CEPC baseline detector}",
    doi = "10.1140/epjp/s13360-020-00272-4",
    journal = "Eur. Phys. J. Plus",
    volume = "135",
    number = "3",
    pages = "274",
    year = "2020"
}

@article{Ruan_ArborPerf_2018,
    author = "Ruan, Manqi and others",
    title = "{Reconstruction of physics objects at the Circular Electron Positron Collider with Arbor}",
    eprint = "1806.04879",
    archivePrefix = "arXiv",
    primaryClass = "hep-ex",
    doi = "10.1140/epjc/s10052-018-5876-z",
    journal = "Eur. Phys. J. C",
    volume = "78",
    number = "5",
    pages = "426",
    year = "2018"
}

@article{PDG,
    author = "Workman, R. L. and Others",
    collaboration = "Particle Data Group",
    title = "{Review of Particle Physics}",
    doi = "10.1093/ptep/ptac097",
    journal = "PTEP",
    volume = "2022",
    pages = "083C01",
    year = "2022"
}

@article{Qu:2019gqs,
    author = "Qu, Huilin and Gouskos, Loukas",
    title = "{ParticleNet: jet tagging via particle clouds}",
    eprint = "1902.08570",
    archivePrefix = "arXiv",
    primaryClass = "hep-ph",
    doi = "https://doi.org/10.1103/PhysRevD.101.056019",
    journal = "Phys. Rev. D",
    volume = "101",
    number = "5",
    pages = "056019",
    year = "2020"
}

@article{delphes,
    author = "Chen, Cheng and Mo, Xin and Selvaggi, Michele and Li, Qiang and Li, Gang and Ruan, Manqi and Lou, Xinchou",
    title = "{Fast simulation of the CEPC detector with Delphes}",
    eprint = "1712.09517",
    archivePrefix = "arXiv",
    primaryClass = "hep-ex",
    month = "12",
    year = "2017"
}

@article{Suehara:2015ura,
    author = "Suehara, Taikan and Tanabe, Tomohiko",
    title = "{LCFIPlus: A Framework for Jet Analysis in Linear Collider Studies}",
    doi = "https://doi.org/10.1016/j.nima.2015.11.054",
    journal = "Nucl. Instrum. Meth. A",
    volume = "808",
    pages = "109--116",
    year = "2016"
}

@article{Catani:1991hj,
    author = "Catani, S. and Dokshitzer, Yuri L. and Olsson, M. and Turnock, G. and Webber, B. R.",
    title = "{New clustering algorithm for multijet cross-sections in $e^+ e^-$ annihilation}",
    reportNumber = "CAVENDISH-HEP-91-5",
    doi = "https://doi.org/10.1016/0370-2693(91)90196-W",
    journal = "Phys. Lett. B",
    volume = "269",
    pages = "432--438",
    year = "1991"
}

@article{CEPC_TDR_Acc,
    author = "Abdallah, Waleed and others",
    collaboration = "CEPC Study Group",
    title = "{CEPC Technical Design Report: Accelerator}",
    eprint = "2312.14363",
    archivePrefix = "arXiv",
    primaryClass = "physics.acc-ph",
    reportNumber = "IHEP-CEPC-DR-2023-01, IHEP-AC-2023-01",
    doi = "10.1007/s41605-024-00463-y",
    journal = "Radiat. Detect. Technol. Methods",
    volume = "8",
    number = "1",
    pages = "1--1105",
    year = "2024"
}

@inproceedings{refTDR_report,
  author       = {Chenguang Zhang},
  title        = {{Global Physics Performance at the Ref-TDR}},
  year         = {2025},
  month        = jun,
  day          = {17},
  url          = {https://indico.ifae.es/event/2054/contributions/9525/attachments/2903/4524/20250617.pdf},
  urldate      = {2025-08-04}
}

@book{ECFA,
    author = "Altmann, J. and others",
    title = "{ECFA Higgs, electroweak, and top Factory Study}",
    eprint = "2506.15390",
    archivePrefix = "arXiv",
    primaryClass = "hep-ex",
    reportNumber = "CERN-2025-005",
    doi = "10.23731/CYRM-2025-005",
    isbn = "978-92-9083-700-8, 978-92-9083-701-5",
    series = "CERN Yellow Reports: Monographs",
    volume = "5/2025",
    month = "6",
    year = "2025"
}

@article{Zhu:2025eoe,
    author = "Zhu, Yongfeng and Liang, Hao and Wang, Yuexin and Che, Yuzhi and Wang, Hengyu and Zhou, Chen and Qu, Huilin and Ruan, Manqi",
    title = "{Holistic approach and Advanced Color Singlet Identification for physics measurements at high energy frontier}",
    eprint = "2506.11783",
    archivePrefix = "arXiv",
    primaryClass = "hep-ex",
    month = "6",
    year = "2025"
}

@inproceedings{ECFA_DRD,
  author    = {Thomas Bergauer},
  title     = {From ECFA Detector Roadmap to 
DRD Collaborations and beyond},
  year      = {2025},
  month     = {June},
  url       = "https://agenda.infn.it/event/44943/contributions/263377/attachments/137562/207083/2025-06-ESPP-Venice-Bergauer\_v4b.pdf",
  note      = "Presentation at ESPP Venice Workshop, INFN, 2025"
}

@article{ILC_TDR,
    author = "Abramowicz, Halina and others",
    title = "{The International Linear Collider Technical Design Report - Volume 4: Detectors}",
    eprint = "1306.6329",
    archivePrefix = "arXiv",
    primaryClass = "physics.ins-det",
    reportNumber = "ILC-REPORT-2013-040, ANL-HEP-TR-13-20, BNL-100603-2013-IR, IRFU-13-59, CERN-ATS-2013-037, COCKCROFT-13-10, CLNS-13-2085, DESY-13-062, FERMILAB-TM-2554, IHEP-AC-ILC-2013-001, INFN-13-04-LNF, JAI-2013-001, JINR-E9-2013-35, JLAB-R-2013-01, KEK-REPORT-2013-1, KNU-CHEP-ILC-2013-1, LLNL-TR-635539, SLAC-R-1004, ILC-HIGRADE-REPORT-2013-003",
    month = "6",
    year = "2013"
}

@article{CLIC_CDR,
    author = "Abusleme Hoffman, A. C. and others",
    editor = {Dannheim, D. and Kr{\"u}ger, K. and Levy, A. and N{\"u}rnberg, A. and Sicking, E.},
    title = "{Detector Technologies for CLIC}",
    eprint = "1905.02520",
    archivePrefix = "arXiv",
    primaryClass = "physics.ins-det",
    reportNumber = "CERN-2019-001",
    doi = "10.23731/CYRM-2019-001",
    volume = "1/2019",
    month = "5",
    year = "2019"
}

@article{FCCee_CDR,
    author = "Barchetta, Nicola and Collins, Paula and Riedler, Petra",
    title = "{Tracking and vertex detectors at FCC-ee}",
    eprint = "2112.13019",
    archivePrefix = "arXiv",
    primaryClass = "physics.ins-det",
    doi = "10.1140/epjp/s13360-021-02323-w",
    journal = "Eur. Phys. J. Plus",
    volume = "137",
    number = "2",
    pages = "231",
    year = "2022"
}

@misc{CEPC2025TDR,
      title={CEPC Technical Design Report -- Reference Detector}, 
      author={The CEPC Study Group},
      year={2025},
      eprint={2510.05260},
      archivePrefix={arXiv},
      primaryClass={hep-ex},
      url={https://arxiv.org/abs/2510.05260}, 
}

@article{Zhu2023PID,
  title   = {Requirement analysis for $dE/dx$ measurement and particle identification performance at the CEPC baseline detector},
  author  = {Zhu, Y. and others},
  journal = {Nucl. Instrum. Meth. A},
  volume  = {1047},
  pages   = {167835},
  year    = {2023},
  doi     = {10.1016/j.nima.2022.167835}
}

@article{Yu2025PID,
  author  = {Yu, Dian and Ding, Houqian and Fan, Yunyun and Zhu, Yongfeng and Qi, Ming},
  title   = {Evaluation of particle identification performance at CEPC with combined $dE/dx$ and time-of-flight information},
  journal = {Nucl. Instrum. Meth. A},
  year    = {2025},
  doi     = {10.1016/j.nima.2025.169163}
}

\end{document}